\documentclass[letterpaper,prl,twocolumn,floatfix,superscriptaddress]{revtex4-1}  %double space preprint format
\usepackage{times}

\usepackage[normalem]{ulem}  %for underlines and strike throughs 
\usepackage{amsmath}
\usepackage[mathcal]{euscript}
\usepackage{mathrsfs}
\usepackage{float}
\usepackage[usenames]{color} %for colored text
\usepackage{graphicx}     %use this if you're using pdflatex to include hyperlinks
\usepackage{hyperref}           %include this if you're using hyperlinks

\hypersetup{
  pdfauthor = {Jacob Stevenson and Peter Wolynes}, 
  pdftitle = { The Ultimate Fate of Supercooled Liquids }, 
  bookmarks=true,         % show bookmarks bar?
    unicode=false,          % non-Latin characters in Acrobat’s bookmarks
    pdftoolbar=true,        % show Acrobat’s toolbar?
    pdfmenubar=true,        % show Acrobat’s menu?
    pdffitwindow=false,     % window fit to page when opened
    pdfstartview={FitH},    % fits the width of the page to the window
    pdfsubject={Subject},   % subject of the document
    pdfcreator={Creator},   % creator of the document
    pdfproducer={Producer}, % producer of the document
    pdfkeywords={keywords}, % list of keywords
    pdfnewwindow=true,      % links in new window
    colorlinks=false,       % false: boxed links; true: colored links
    linkcolor=red,          % color of internal links
    citecolor=green,        % color of links to bibliography
    filecolor=magenta,      % color of file links
    urlcolor=cyan           % color of external links
}

\newcommand{\na}{n_{\alpha}}
\newcommand{\nn}{n_{\nu}}
\newcommand{\nMdag}{n_{M}^{\ddagger}}
\newcommand{\xia}{\xi_{\alpha}}
\newcommand{\xiM}{\xi_{M}}

\newcommand{\ratea}{\omega_{\alpha}}
\newcommand{\raten}{\omega_{\nu}}
\newcommand{\rateM}{\omega_{M}}
\newcommand{\ratens}{\omega_{M\nu}}
\newcommand{\siga}{\sigma_{\alpha}}
\newcommand{\sigx}{\sigma_{x}}
\newcommand{\ya}{y_{\alpha}}
\newcommand{\yx}{y_{x}}
\newcommand{\deltae}{\Delta \epsilon}
\newcommand{\Tcross}{T^{\chi}_{\mathrm{cross}}}
\newcommand{\Tcrit}{T_{\mathrm{crit}}}

\newcommand{\sccrit}{s_c(\Tcrit)}

\newcommand{\nnhetero}{{\nn^{\mathrm{hetero}}}}
\newcommand{\phetero}{p_{\mathrm{hetero}}}

\newcommand{\ratenshetero}{{\ratens^{\mathrm{hetero}}}}

\newcommand{\figsizeplot}{\ref{fig:size_plot}}
\newcommand{\figrelaxationtimes}{\ref{fig:relaxation_times}\textbf{A}}

\makeatletter

\newcommand{\Rmnum}[1]{\expandafter\@slowromancap\romannumeral #1@}
\makeatother

\begin{document}

\title{ The Ultimate Fate of Supercooled Liquids }
\date{\today}
\author{Jacob D. Stevenson}
\affiliation{Department of Physics and Department of Chemistry and Biochemistry,
Center for Theoretical Biological Physics,
University of California at San Diego, La Jolla, California 92093}
\affiliation{Institut f\"ur Physik, 
Johannes Gutenberg-Universit\"at, 55099 Mainz, Germany}
\author{Peter G. Wolynes}
\affiliation{Department of Physics and Department of Chemistry and Biochemistry,
Center for Theoretical Biological Physics,
University of California at San Diego, La Jolla, California 92093}
\affiliation{e-mail: pwolynes@ucsd.edu}

\begin{abstract}

In recent years it has become widely accepted that a dynamical length scale
$\xia$ plays an important role in supercooled liquids near the glass
transition.  We examine the implications of the interplay between the growing
$\xia$ and the size of the crystal nucleus, $\xiM$, which shrinks on cooling.
We argue that at low temperatures where $\xia > \xiM$ a new crystallization
mechanism emerges enabling rapid development of a large scale web of sparsely
connected crystallinity.  Though we predict this web percolates the system at
too low a temperature to be easily seen in the laboratory, there are noticeable
residual effects near the glass transition that can account for several
previously observed unexplained phenomena of deeply supercooled liquids
including Fischer clusters, and anomalous crystal growth near $T_g$.

\end{abstract}

\maketitle

%\tableofcontents
\bibliographystyle{naturemagurl}

Thermodynamics tells us that the ultimate fate of a liquid held below its
melting point is to crystallize. This notion is buttressed by the observation
that terrestrial rocks are generally polycrystalline and the rare amorphous
minerals found naturally are geologically young\cite{rogers.1917}.
Nevertheless, amorphous solids are ubiquitous and while some atactic polymers
or heteropolymers may not be able to crystallize at all because they have no
plausible competing periodic crystal structure, most everyday glass substances
are only kinetically prevented from crystallizing on human time scales.
Understanding the competition between crystallization and glass formation is
thus of great practical significance. Turnbull's early ideas about this
competition, based on augmenting nucleation theory with a dynamical correction
from the viscous slowing of glassy liquids\cite{greet.1967}, have held up
remarkably well, allowing the discovery and exploitation of metallic
glasses\cite{chaudhari.1978,duwez.1967}, among other things. Reasoning
analogous to Turnbull's inspired the energy landscape theory of protein
folding, leading to the idea that proteins have evolved to avoid the kinetic
traps expected for heteropolymers, allowing rapid formation of native
structure\cite{bryngelson.1995}: the ``minimal frustration principle.''
Turnbull's nucleation argument also implies a crisp time scale separation
between crystallization and the quiescent equilibrium dynamics of a supercooled
liquid.  This time scale separation makes it possible to discuss an equilibrium
supercooled liquid as defined by the Gibbs measure applied to that part of many
body configuration space lacking supercritical crystallization nuclei. This
restricted equilibrium description would be useful down to a temperature where
thermodynamically barrierless or spinodal crystallization can occur.

Many features of the dynamics of metastable, ``equilibrated'' supercooled
liquids, and of nonequilibrium glasses, have been understood using the random
first order transition (RFOT) theory of the glass
transition\cite{lubchenko.2007}. Generally, it has been possible, within the
RFOT framework, to ignore the possibility of a periodic crystalline state. In
this paper we explore how RFOT theory is modified when we account for the
existence of a periodic crystalline state. These arguments suggest there is
indeed a wide range of accessible thermodynamic conditions where Turnbull's
analysis of nucleation should hold. These arguments predict, however, that even
above the temperature where there is a strict spinodal, a new mechanism of
crystallization should emerge that will compromise the usually assumed time
scale separation between crystallization and structural (so called $\alpha$)
relaxation. When a substance has a periodic ground state, an approach to an
ideal glass transition amongst the strictly amorphous structures is then
predicted to ultimately be intercepted by crystallization in a kinetic sense. 

The arguments in this paper hinge on the dynamical mosaic structure envisioned
in RFOT theory. RFOT theory predicts the existence of a length scale for
dynamical correlations that grows as the liquid is cooled. Long after this
length scale was predicted, numerous experiments using NMR and imaging directly
revealed such a length scale for dynamical
heterogeneity\cite{ediger.2000,russell.2000,mackowiak.2009}. The RFOT predicted
length agrees with both those direct measurements and more indirect inferences
of dynamical length scales\cite{berthier.2005,capaccioli.2008}.  Since the
dynamical correlation length grows on cooling, the scales of the dynamical
mosaic at some point become comparable to the size of the classical
crystallization nucleus which, in contrast, shrinks as the liquid becomes more
deeply supercooled. Just below freezing, when the dynamical and nucleation
length scales are well separated, the Turnbull nucleation picture is adequate.
But upon deeper supercooling, when the scales cross, a new kind of local
percolative ``nanocrystallization'' is predicted to occur, driven by the
dynamical heterogeneity of the viscous liquid. At first, the nuclei will be
sparse and grow slowly at a rate still controlled by $\alpha$ relaxation. A bit
further cooling, however, allows another threshold to be crossed where the
theory predicts more extensive crystal nucleation and considerably more rapid
crystal growth, which while still slow in human terms, will be completely
decoupled in time scale from $\alpha$ relaxation.  In this regime a true
$\alpha$ relaxation time for an equilibrium supercooled liquid would not be
operationally defined.

These new regimes of crystallization explain several anomalous observations
about supercooled liquids which suggest that supercooled liquids have bigger
density variations than would be expected on thermodynamic grounds. One such
anomaly is the excess low angle x-ray scattering observed by Fischer's group
indicating fluctuations whose magnitude is inconsistent with the macroscopic
compressibility\cite{fischer.1993}. Our picture ascribes this excess
scattering to slow growing nanocrystallites. The nanocrystallites also explain
the long environmental exchange times observed in some single molecule
fluorescence experiments\cite{zondervan.2007}, that have so far
not been observed using nonlinear NMR\cite{mackowiak.2009}.  If sufficient time
elapses, the nanocrystallites can grow to form a connected but fragile
web\cite{zondervan.2007}, explaining anomalous solid-like
elastic behavior at low shear, (which disappears on more vigorous flow) even
above the nominal glass transition temperature. In this new regime,
heterogeneous nucleation from the presence of foreign or seeded nuclei is also
enhanced, explaining the quantitative inconsistencies between the various
experiments that find such anomalous long-lived structures.

The present picture also predicts that crystal growth starting from
heterogeneously nucleated crystals will abruptly change speed and mechanism
upon sufficiently deep supercooling. Several experiments do show
anomalously rapid crystal growth beginning near, but above, the laboratory glass
transition\cite{hikima.1999,hikima.1995,sun.2008}.

The organization of the paper is as follows: we first reprise the ``energy
landscape library construction''\cite{lubchenko.2004}, now accounting for the
gap in the spectrum of states that arises from the existence of a particularly
stable periodic structure. This argument defines the length scales for motions
relevant to $\alpha$ relaxation, the classical crystallization nucleus size and
a third critical size for forming nanocrystallites. The comparisons of these
lengths allow us to delineate the crystallization regimes.  Similar arguments
for heterogeneous nucleation on foreign nuclei makes predictions for crystal
growth which we compare to experiment.  Finally, we discuss more general issues
of phase separation and polyamorphism in glassy liquids as well as how these new
mechanisms impinge on the design of glassy materials and on protein folding
theory.

\section{   Energy Landscape Libraries and Crystallization}

\begin{figure}
\centering
\includegraphics{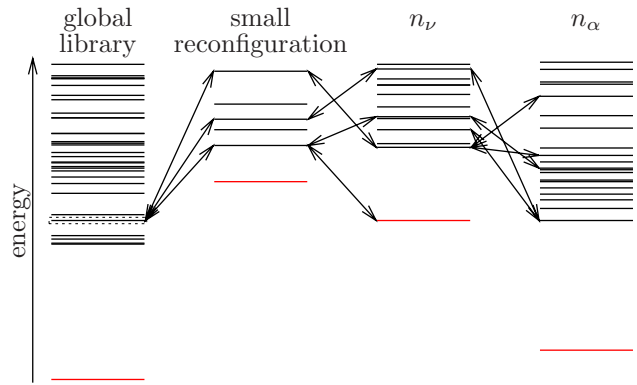}
\caption{A schematic diagram of the microcanonical energy landscape. The
leftmost panel shows the global configuration of states with the crystal having
the lowest energy.  The second panel shows the distribution of
energy levels after a small number of particles have reconfigured (starting
from the boxed initial state).  The energies are higher due to the surface
mismatch penalty. The third panel shows the distribution of energies after
$\nn$ particles have moved.  Here the energy of the crystal nucleus is
comparable to the initial state, but is destabilized by the entropy of the
large number of amorphous configurations.  The rightmost panel shows the
distribution of energies after reconfiguring $\na$ particles.  The arrows suggest possible
pathways of reconfiguration.
}
\label{fig:library_construction}
\end{figure}
%At this point an
%amorphous configuration exists which has comparable energy to the initial state
%making stable amorphous reconfiguration possible.  

A liquid has a large diversity of structures, as manifested by an
extensive configurational entropy. In figure \ref{fig:library_construction}, we
show schematically the spectrum of local free energy minima of a large sample.
Macroscopic crystalline states have an extensive gap
separating them, in energy, from the liquid manifold.  Since the thermally
sampled states change with temperature, this gap is taken to be $\Delta
\epsilon (T) N$ where $N$ is the number of molecular units in the sample. At
the melting point the magnitude of $\Delta \epsilon$ is accessible from
calorimetry $\Delta \epsilon(T_m) = T_m S_c(T_m)$. For simplicity, we will
generally speak of an energy landscape library, although more properly we
should use the term (Gibbs) free energy landscape since local vibrational
entropies should be included --- indeed, for hard spheres, vibrational entropy
provides the entirety of an individual minimum's free energy. 

Structural transitions occur through local rearrangements. Thus we must
consider the changes of (free) energy and diversity that occur when only a
finite number of particles, $n$, are moved. Instead of a single global energy
landscape, there are thus numerous local energy landscape libraries as
illustrated also in Fig \ref{fig:library_construction}. In general, when a
small number of molecules move, the energy becomes higher than the initial
minimum's energy since the alternate geometry has steric conflicts with the
original surroundings, now frozen, which can only be partially ameliorated by
small harmonic deformations.  Owing to the extensive configurational entropy,
the size of the local library grows exponentially with the number of displaced
particles $n$, and at some point, with increasing $n$, a structure of
comparable energy to the original one will be found. If the energy gap is
substantial, the first near resonant level to be encountered will be
crystalline. We will call such a structure a ``nanocrystallite.'' It will
contain $\nn$ particles. Although such a nanocrystallite can initially form
locally, if the configurational entropy is high the newly formed
nanocrystallite will quickly disappear, because although transitions to any of
the other specific amorphous states is energetically uphill, such transitions
are extraordinarily numerous --- roughly there are $\exp \{\nn s_c / k_B\}$  of
them, giving their net formation rate a large value.

Owing to its entropic disadvantage, a nanocrystallite, after forming, will
revert to one of the set of amorphous members of the local library.  It will
then usually fall back downhill to the original amorphous structure since $n$
is less than $\na$.  Of course, a distinct amorphous structure could become
near resonant, and indeed this typically happens when $\na$ particles are
displaced, defining the size of a cooperatively rearranging region (of radius
$\xi_{\alpha}$). Eventually, as bulk thermodynamics dictates, the entropic
disadvantage of a small nanocrystallite will be overcome by the growth with
size of the energy gap, so a big enough nanocrystallite will not disappear, but
grows, essentially irreversibly.  This point of no return defines the classical
critical nucleus for crystallization, $\nMdag$.

There are three different free energy curves to consider when thinking about
local rearrangements of a supercooled liquid with a periodic crystalline ground
state. These can be thought of as reflecting separate averages over the
amorphous states and the periodic crystal states treated as reactants and
products in a chemical reaction. (At this stage we will ignore the fluctuations
in driving force that will modify these free energy profiles owing to the
diversity of surrounding environments. Let us take, for concreteness, the
mismatch energy of amorphous/amorphous pairings to scale like $\siga n^{\ya}$
and the mismatch energy for the amorphous/crystal pairings to scale as $\sigx
n^{\yx}$.  Such pure power laws are oversimplifications. For compact
reconfiguring clusters in three dimensions, the exponents in these expressions
would be 2/3.  RFOT theory suggests there are significant ``wetting''
corrections to these (mean field) results; these corrections may be large
enough (near an ideal glass transition) to give an effective exponent of 1/2
rather than 2/3. While the most appropriate exponent is
debated\cite{stevenson.2008a,stevenson.2006,capaccioli.2008,biroli.2008,tracht.1998},
the distinction between various choices makes little difference to the present
story, as we shall see. Likewise the crystallization mismatch energy exponent
$\yx$ would classically be expected to be 2/3, but could appear smaller if the
crystallite's surface is above its a roughening transition.

The free energy for a reconfiguration event to any other amorphous state
satisfies $F_{\alpha} = -Ts_c(T) n + \siga n^{\ya}$.  The reader is referred to
the paper by Lubchenko and Wolynes\cite{lubchenko.2004} to see how the
microcanonical rate of conversion to any \emph{one} specific amorphous state is
transformed to the rate for transiting to the ensemble of amorphous states
(hence the driving force is $-T s_c(T)$).  The profile for transiting directly
to a nanocrystallite (since a single definite starting state is involved) has
the energy gap driving force, hence $F_{\nu} = -\Delta \epsilon(T) n + \sigx
n^{\yx}$. Yet, if the numerous alternate amorphous structures can also be
accessed by the disappearance of transient nanocrystallites and $\alpha$
relaxation processes, the bulk driving force will be that from an equilibrated
ensemble of amorphous structures, i.e. per particle it is $-( \Delta
\epsilon(T) - Ts_c(T)) $, giving the total macroscopic nucleation free energy
profile $F_M = -( \Delta \epsilon(T) - Ts_c(T)) n + \sigx n^{\yx}$.

Each of the three curves defines characteristic sizes: $\nMdag$, the size of
the macroscopic crystal transition state nucleus, $\na$ the size of a
cooperatively rearranging amorphous region and $\nn$ the size of a
nanocrystallite.  Their values are found from the free energy profiles
described above:

\begin{eqnarray}
\nonumber
\na &=& \left( \frac{\siga}{ T s_c(T) } \right)^{\frac{1}{1-\ya}} \\
\nn &=& \left( \frac{\sigx}{ \deltae(T) } \right)^{\frac{1}{1-\yx}} 
\label{eqn:sizes}
\\
\nonumber
\nMdag &=& \left( \frac{\sigx \yx}{ \deltae(T) - T s_c(T) } \right)^{\frac{1}{1-\yx}}
.
\end{eqnarray}

\begin{figure}
\centering
\includegraphics{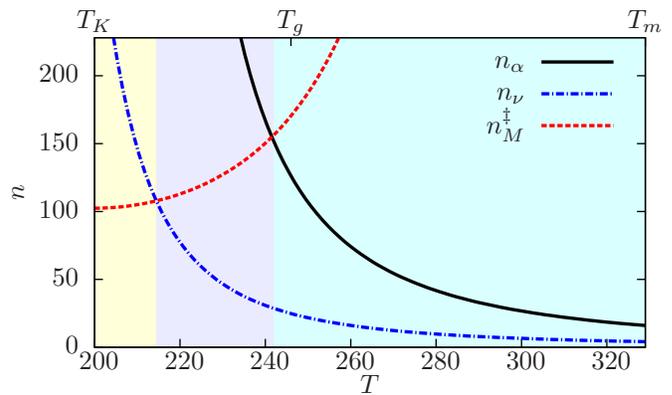} 
\caption{The temperature dependence of $\na$, the size of a
typical amorphous reconfiguration;  $\nMdag$, the transition state size of
classical crystal nucleation; and $\nn$, the number of particles involved in
nanocrystallization.  Within the shaded region on the right classical
nucleation theory is valid.  In the shaded region on the left direct
nanocrystallization can take place.  Crystallization in the center region
takes place through fluctuational, percolative nanocrystallization.  
}
\label{fig:size_plot}
\end{figure}

\begin{figure}
\centering
\begin{tabular}{p{\textwidth}}
%\textbf{A} \\ [-2ex]
%\input{relaxation_times} \\
\includegraphics{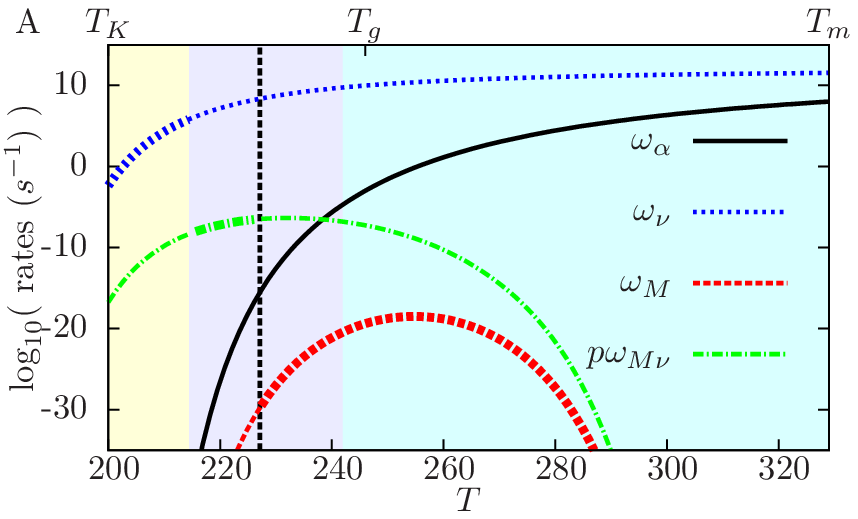} \\
%\textbf{B} \\ [-2ex]
%\input{heterogeneous_growth} 
\includegraphics{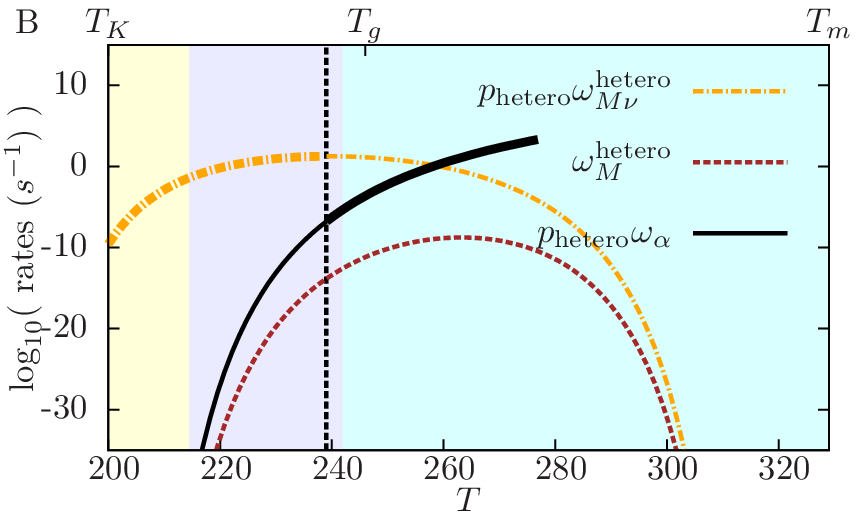} 
\end{tabular}
\caption{Panel \textbf{A} shows the temperature dependence of nucleation rates for
$\alpha$ relaxation ($\ratea$), bulk crystallization ($\rateM$),
nanocrystallization ($\raten$), and percolative fluctuational
nanocrystallization ($p \ratens$). The temperature regimes are shaded as per
figure \protect{\ref{fig:size_plot}}. The homogeneous percolation transition is
marked with a vertical line.
Rates are indicated as the inverse time to see a nucleation event within a
microscopic volume. Since a single crystal nucleus would spread rapidly through
the system the rate at which a sample of volume $V$ would crystallize is $\sim
\rateM V$, a rate much faster than that appearing in the figure.
Panel \textbf{B} shows rates for heterogeneous crystal surface nucleation.  At
high temperatures the surface nucleation rate is proportional to $\ratea$.  At
the heterogeneous percolation temperature (marked by a vertical line) the
crystal growth rate switches to be $\phetero \ratenshetero$. For both panels
the relevant rates for crystal nucleation and
growth are emphasized with a thick line.
}
\label{fig:relaxation_times}
\label{fig:heterogeneous_growth}
\end{figure}

\noindent As sketched in figure \figsizeplot, $\na$ increases with decreasing $T$,
and in contrast $\nMdag$ decreases with decreasing $T$.  The nanocrystallite size
changes the least with temperature.  Figures \ref{fig:size_plot} and
\ref{fig:relaxation_times} are drawn using material parameters of the fragile
liquid o-terphenyl, the details of which, as well as analogous curves for a
strong glass former, can be found in the supplementary material.

\begin{figure}
\centering
\includegraphics{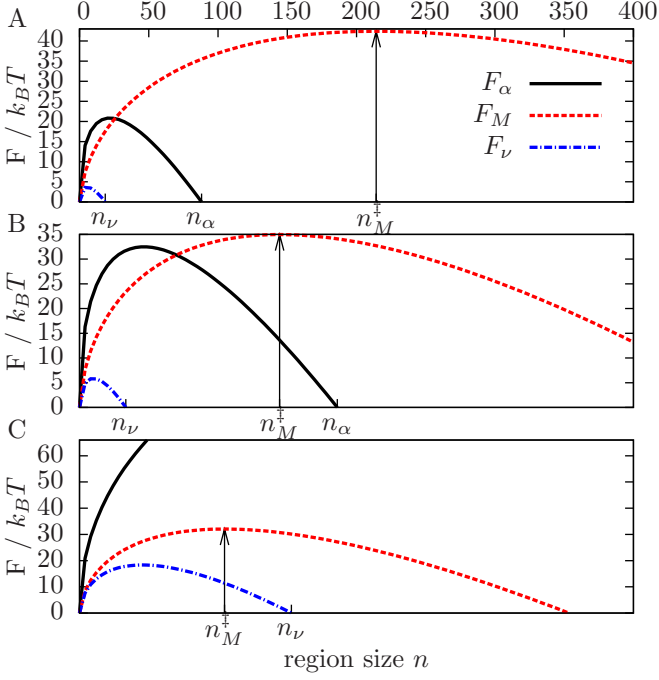}
\caption{ Free energy profiles of $\alpha$ relaxation ($F_{\alpha}$), bulk
crystallization ($F_M$) and nanocrystallization ($F_{\nu}$) at the three
temperature regimes shaded in figure
\protect{\figsizeplot}. }
\label{fig:FE_curves}
\end{figure}

Macroscopic nucleation only occurs at temperatures below the melting point
$T_m$.  Just below $T_m$ the relationships between the three free energy
profiles are as pictured in figure \ref{fig:FE_curves}a. In this regime $\nMdag
> \na > \nn$.  Nanocrystallites constantly form and disappear either reverting
to the initial amorphous state or becoming new amorphous structures in which
$\na$ particles have moved. The macroscopic nucleation barrier at $\nMdag$ is
crossed by making many moves of size $\na$. Thus the prefactor of the rate will
be related to $\ratea$.  In this high temperature regime the rates associated
with each of the processes discussed are 

\begin{eqnarray}
\nonumber
\ratea &=&
\omega_0 \exp \left\{ - \frac{\siga}{k_B T}  \left( \frac{\ya \siga}{Ts_c(T)}
\right)^{\frac{\ya}{1-\ya}} (1-\ya) \right\} \\
\raten &=&
\omega_0 \exp \left\{ - \frac{\sigx}{k_B T}  \left( \frac{\yx \sigx}{\deltae(T)}
\right)^{\frac{\yx}{1-\yx}} (1-\yx) \right\} 
\label{eqn:relaxation_times}
\\
\nonumber
\rateM &=&
\ratea \exp \left\{ - \frac{\sigx}{k_B T}   \left( \frac{\yx \sigx}{ \deltae(T) - Ts_c(T) }
\right)^{\frac{\yx}{1-\yx}} (1-\yx) \right\}
\end{eqnarray}

\noindent The prefactor for both $\raten$ and $\ratea$ is $\omega_0$, the rate
of microscopic vibrations.  The temperature dependence of the rates is shown in
figure \figrelaxationtimes.  It should be noted that for the bulk
crystallization rate the prefactor ought to be the average inverse $\alpha$
relaxation time, however for simplicity, we use the inverse typical
relaxation time, underestimating the bulk crystallization rate, especially
for lower temperatures.

The mechanism of nucleation changes when the supercooling is sufficient for
$\na$ to exceed $\nMdag$, as shown in figure \ref{fig:FE_curves}b and figure
\ref{fig:FE_curves}c, i.e. at $\Tcross$, defined by

\begin{equation}
\na(\Tcross) \equiv \nMdag(\Tcross).
\end{equation}

\noindent It can be seen from figure \figsizeplot~ that this crossover occurs
near the laboratory $T_g$.  This concurrence is fairly robust to the details of
the liquid, but is a coincidence, since the one hour time scale used for the
laboratory glass transition is anthropocentrically defined.

There are two somewhat different mechanisms of crystallization below the
temperature $\Tcross$ where the traditional nucleation theory analysis breaks
down. Which mechanism applies depends on the relative size of $\nn$ and
$\nMdag$.  Only at very low configurational entropies can we satistfy the
strongest condition $\nn > \nMdag$. In this case, shown in figure
\ref{fig:FE_curves}c the nanocrystallite has no tendency to disappear at all.
Once a nanocrystallite forms, it continues to grow without having a chance to
access other amorphous structures.  The critical temperature for in this
extremely supercooled regime $\nn(\Tcrit) \equiv \nMdag(\Tcrit)$ leads, through
equation \ref{eqn:sizes}, to the relation

\begin{eqnarray}
 \Tcrit \sccrit =
(1-\yx) \deltae(\Tcrit) .  
\end{eqnarray}

\noindent It can be seen in figure \figsizeplot~ that this
direct form of nanocrystallization occurs significantly below $T_g$, at deep
enough supercooling that it probably has not been observed in the laboratory,
although it may occur in geology.  

When crystals form from direct nanocrystallization rather than through the
rearrangements of $\alpha$ relaxation, the prefactor for the nucleation rate
will be $\raten$ rather than the much slower $\ratea$

\begin{equation}
\ratens = \raten \exp \left\{ -\frac{F^{\ddagger}_M(T)}{k_B T} \right\}
\end{equation}

\noindent Were it possible to reach this regime of undercooling while remaining
equilibrated in the amorphous ensemble, there would clearly be no dependence of
crystallization rate on $\ratea$. The sample would completely crystallize at the
rate $\ratens$ before any $\alpha$ relaxation time could be measured.

\section{The dynamical mosaic and percolating nanocrystallites}

There is an intermediate regime not covered by the above cases. What happens
when $\na > \nMdag$ but $\nMdag > \nn$ as shown in figure \ref{fig:FE_curves}b?
No single process dominates, so the answer depends on an interplay between
length and time scales for the fluctuations.  Typically a nanocrystallite, once
formed, disappears, but no alternate amorphous structure can be stabilized at
that location.  Nevertheless since $\nMdag$ is smaller than $\na$, the
nucleation threshold could still be crossed while remaining within the original
local amorphous structure.  According to RFOT theory, each region of size $\na$
can be thought of as having a variable but temporarily fixed
energy\cite{xia.2001,stevenson.2008a} (until its environment reconfigures). The
magnitude of the energy variance of a region of size $n$ follows from mesoscale
statistical thermodynamics and is related to the configurational heat capacity,
$T^2 k_B \Delta C_p / n$.

In some mosaic cells, the driving force will be sufficient for the
nanocrystallite to cross the macroscopic crystal formation threshold. In those
cells then, nanocrystallites will initially form. The density
of these stable nanocrystallites will depend on $\sccrit$, the critical entropy
density for direct nanocrystallization that we have already discussed.  We
expect a fraction of mosaic cells 

\begin{equation}
p = \frac{1}{2} \mathrm{erfc} \left\{ \frac{s_c(T)
- \sccrit}{\sqrt{2 k_B \Delta C_p / \nn } } \right\} 
\end{equation}

\noindent will irreversibly nucleate in this manner.  The temperature
dependence of $p$ is displayed in the supplementary material. If this
nanocrystallization probability is big enough to allow percolation, a gossamer
percolative network will form rapidly at the rate $p \ratens$.  This has the
effect of raising the lower critical temperature which signals the onset of
rapid crystallization independent of the slow $\alpha$ relaxations.  Even above
this percolation transition temperature, if $\alpha$ relaxation is slow
compared with the nanocrystallite nucleation rate $p \ratens$, large, ramified
networks of crystalline structure will appear, but remain finite in size.  The
details of crystal morphology in this regime involve a complicated interplay of
length and time scales and deserve a more extensive treatment than is provided
here. At high temperatures the formation of these crystalline networks will be
broken up by amorphous reconfigurations, destroying any nascent long range
order. 

\section{Heterogeneous Crystal Nucleation and Growth on preexisting Crystals}

Any of the sparse, stable nanocrystallites, once formed, will be able to grow
directly, in the manner of a chain reaction, if it has a further neighbor that
is itself also sufficiently stable. Each step in a possible nucleation chain is
actually a heterogeneous nucleation process. The presence of a nanocrystallite
increases the rate at which its neighbors nucleate.  Foreign particles or seed
nuclei likewise make possible smaller critical nucleation sizes, and thus the
percolative crystallization mechanism can begin at lower degrees of super
cooling when seeds are present. If the seeds introduced are nanoscale,
smaller than$\na$ in size, they can be thought
of as adding a constant stability increment to the crystallization free energy
profiles, increasing locally the driving force for nucleation, and allowing the
nanocrystallite chain reaction route at a higher temperature than for strictly
homogeneous nucleation. 

The situation for still larger seed particles, those bigger than $\na$, is
somewhat more subtle. For nucleated crystal growth on a flat interface only a
hemisphere of newly crystallized material need be laid down at a
time\cite{hikima.1995,stevenson.2008b}. The disturbed volume is approximately
half the volume for spherical growth $\nnhetero \approx \nn/2$.  The
propagation probability

\begin{equation}
\phetero = \frac{1}{2} \mathrm{erfc} \left\{
\frac{s_c(T) - \sccrit}{\sqrt{2 k_B \Delta C_V / \nnhetero } } \right\}, 
\end{equation}

\noindent is greater than $p$, since growth after the first nucleus is present
is essentially a surface process.  The degree of stabilization for growth on a
curved surface will be somewhat smaller.  A more general analysis would take
$\nn/2 \leq \nnhetero \leq \nn$, but we restrict our discussion here to the
limiting case.  The rate for this heterogeneous percolative growth should be of
order $\phetero \ratenshetero$ where

\begin{equation}
 \ratenshetero \approx 
\raten
\exp \left\{ \frac{-F^{\ddagger}_{M}(T)}{ 2 k_B T} \right\}
\end{equation}

\noindent which is much faster than $\ratea$ at low temperatures.  The chain
reaction mechanism ultimately turns off as $s_c$ increases.  After falling
below the percolation criterion upon warming, crystal growth will proceed a
finite distance before it is obliged to wait for the environment to
reconfigure.  Thus at high temperatures the growth is limited by the $\alpha$
relaxation rate and will be proportional to $\phetero \ratea$, as in figure
\ref{fig:heterogeneous_growth}.  Since $\phetero(T)$ crosses the percolation
threshold at a higher temperature than $p(T)$, seed nuclei are sparse at this
crossover temperature, leading to a noticeable period of aging during which
$\ratea$ can be measured, but eventually, as always the sample will
crystallize.  An analogy with the percolation interactive cluster growth
model\cite{anderson.1988} would probably yield a more precise description for
the heterogeneous percolation transition.

\section{ Observations of Nanoscale Structures and Anomalous Crystal Growth in
Supercooled Liquids}

Low angle scattering from supercooled liquids in excess of that following from
the compressibility has often been observed, notably by
Fischer\cite{fischer.1993}. Macroscopic crystallinity usually is not
seen during the experiments.  The observations require one to explain how seeds
can have nucleated but not have grown perceptibly. The present analysis does,
in fact, suggest that some nucleation centers appear at temperatures near
$T_g$, owing to fluctuations in driving force, but that these centers still
grow rather slowly at a rate dependent on $\ratea$. While a strictly
homogeneous mechanism may explain the observations, it is likely that foreign
nuclei are involved in some of the experiments, since heterogeneous crystal
growth is also accelerated via percolative nucleation in this regime.  Fischer
carefully noted in his early papers that very pure samples which were only
quenched a single time to the low temperature of investigation, without being
rewarmed, did not exhibit anomalous scattering.  Heterogeneous initiation would
explain the ability to make such so called ``cluster-free'' samples quite
nicely.

Since the nanocrystallites grow by a dynamical percolation or chain reaction
process, the embedded nanocrystallites are expected to be finite initially, but
have a fractal shape. Eventually they will form a web that may be quite
sparsely connected.  In the early stages, particles near the nanocrystals will
rotate more slowly than would be expected in the ``bulk'' supercooled liquid.
Fractal nanocrystallites provide very long-lasting heterogeneities that will
relax on times much larger than $\ratea^{-1}$ which otherwise would be the
natural time scale of environmental renewal in the absence of crystallization.
The large exchange times seen in some single molecule experiments are easily
explained in this way. Predicting the precise magnitude of
$\omega_{\mathrm{exch}} / \ratea$, however, requires further theoretical work
and, more important, better experimental characterization of the preparation
protocol since the size and shape of the nanocrystalline heterogeneities is
preparation time dependent.  It would also be interesting to redo the single
molecule experiments with the protocol Fischer employed for achieving ``cluster
free'' samples.

The precise temperatures at which the new growth mechanism on existing crystals
begin, depends on the crystal amorphous surface energy $\sigx$. This surface
energy, in turn, depends both on the crystal polymorph growing and on the
specific crystal face.  Ediger's observation that anomalous crystal growth
occurs only for some polymorphs is consistent with the present
arguments\cite{sun.2008}. Those forms more prone to anomalously fast growth
should have local structures more consonant with the liquid, and thus smaller
$\sigx$ according to the present theory. This agrees with Ediger's
observations\cite{sun.2008}. 

The percolative character of the growth mechanism tied to $\alpha$ relaxations
should lead to considerable directional randomization on each step. This
results in less dependence of the growth rate on the Miller indices of the
macroscopic crystal than for the normal crystalline growth mechanism, leading
to spherulitic growth.  At $T_g$ percolative nanocrystallite nucleation is
predicted to increase the rate of crystal surface growth by $\ratenshetero /
\ratea \sim 4$ orders of magnitude, in good agreement with the observed speed
up in crystal growth\cite{hikima.1999}.

It is interesting to note that spherulitic growth like that studied in the
laboratory near $T_g$ has also been observed in the geological context. Spheres
of crystalline minerals with fibrous arrangements of crystallites are found
both embedded in volcanic glasses and as free rocks which have sometimes been
(in all likelihood) misinterpreted as unnatural
artifacts\cite{kirkpatrick.1975,smith.2001,stirling.1969}.

\section{Implications}

The dynamical mosaic of RFOT theory has implications for many kinds of phase
transitions in glass forming substances. Our arguments are based merely on the
existence of an energy gap in the local minimum spectrum so they are thus
equally appropriate for any first order transition.  In mixtures phase
separation into components may also occur without or along with
crystallization. If the liquid is quenched deeply into the binodal for
demixing, the analysis of phase separation should be quite parallel to the
present one for crystallization. On the other hand, near a critical point for
phase separation, the length scale of the critical composition fluctuations
enters. How this length (which grows as the critical point is approached)
competes with the RFOT mosaic length scale requires a more elaborate analysis.
Such an analysis could be based on coupling the dynamical Landau-Ginzburg
theory with the fluctuating version of the mobility transport equations
proposed by one of us recently\cite{wolynes.2009}.

The topic of polyamorphism i.e. the existence of multiple noncrystalline
phases, has engendered much spirited
discussion\cite{mishima.1998,wilding.2006}. Some models of the glass transition
based on frustrated phase transitions connect directly to the existence of
polyamorphism\cite{tarjus.2005}. The simplest Hamiltonian model of a frustrated
phase transition, the so-called Brazovskii Hamiltonian for stripe formation,
indeed is predicted to have both a first order polyamorphic transition, \`a la
Brazovskii\cite{brazovskii.1975}, and a mean field random first order
transition, \`a la Schmalian and Wolynes\cite{schmalian.2000}. The present
ideas should illuminate the relations between those two transitions in stripe
glass forming systems.

The present argument also raises some cautionary points about polyamorphism.
First, the ubiquity of nanocrystallite formation at deep undercooling means
that the observation of a nonfacetted, overall isotropic drop of a second phase
in a supercooled liquid may reflect the formation of a disoriented tangle of
nanocrystalline fibers.  Several of the cause c\'el\`ebre substances that have
been thought to exhibit polyamorphism, triphenyl phosphite and butanol for
example, seem to actually be examples of such poly-nanocrystalline tangles: the
Raman vibrational spectra of the spherical inclusions, are identical to
the crystal\cite{wypych.2007}, and the x-ray diffraction of those inclusions
matches closely a linear combination of those for a
nanocrystalline powder and a truly aperiodic isotropic
liquid\cite{shmytko.2010}.

Other examples of polyamorphism such as water, are on stronger footing since
they are inspired by the undoubted existence of distinctly different amorphous
samples prepared by vapor deposition (high and low density amorphous
ices\cite{mishima.1998}) or by pressure-induced
amorphization\cite{mishima.1984}. Yet again, the relation of the new amorphous
phases prepared via strongly nonequilibrium routes to any extrapolated
equilibrium supercooled liquid may well be complicated by the intervention of
the rapidly nucleating crystal phase at low specific configurational entropy.
Indeed, if the extrapolated liquid-liquid transition occurs below the
nanocrystallite formation limit discussed here there may be in principle
difficulties in directly testing such extrapolations.

We see the dynamically corrected nucleation arguments put forward by Turnbull
break down at low temperatures. Yet most of the practical implications of
Turnbull's arguments remain intact under our revision of his crystal nucleation
ideas: a substance with low melting point compared to glass transition will not
only nucleate slowly in the conventional way but will also have a hard time
accessing percolative nanocrystallization. Eutectics remain the best candidates
for glass formation just as Turnbull suggested. In a similar way, protein
folding, when the driving force is strong enough, can kinetically decouple from
trap escape through a percolation mechanism like that discussed
here\cite{bryngelson.1995,shen.2008,naganathan.2007}. Still, the fastest
folders must have evolved to avoid traps, even while it is true that slow
folders eventually will make it to the folded state, anyway. In general the
strongly stabilized, pseudo downhill folding envisioned in this scenario is not
the most common for proteins that have been studied so far in the laboratory.
While strictly downhill folding is not expected to be the dominant mode of
natural protein folding according to current models, this pseudo downhill
mechanism may be more common.

Finally we note that an understanding of the percolative nanocrystallization
kinetics may allow the synthesis of new materials via special heat treatments.
The resulting nano-porcelains may have useful mechanical, electrical and
magnetic properties.

\begin{acknowledgments} 
  Support from NIH grant 5R01GM44557 is gratefully
  acknowledged.  
\end{acknowledgments}

\bibliography{jakes_biblio}

\end{document}